\documentclass[draftcls,onecolumn,11pt]{IEEEtran}

\usepackage{amsmath}

\usepackage{cite}
\usepackage{algorithm}
\usepackage{algpseudocode}
\usepackage{enumerate}
\usepackage{color}
\usepackage{setspace}
\usepackage{balance}
\usepackage{url}
\usepackage{stfloats}
\usepackage{flushend} 

\usepackage[pdftex]{graphicx}
\graphicspath{{./}}%
\DeclareGraphicsExtensions{.jpg,.png,.pdf}
\usepackage[font={footnotesize}]{caption}
\usepackage[caption=false,font=footnotesize]{subfig}

\newcommand{\bi}{\begin{itemize}}	
\newcommand{\ei}{\end{itemize}}
\newcommand{\bn}{\begin{enumerate}}	
\newcommand{\en}{\end{enumerate}}
\newcommand{\bc}{\begin{center}}
\newcommand{\ec}{\end{center}}
\newcommand{\be}{\begin{equation}}
\newcommand{\ee}{\end{equation}}
\newcommand{\bea}{\begin{eqnarray}}
\newcommand{\eea}{\end{eqnarray}}
\newcommand{\ben}{\begin{equation*}}
\newcommand{\een}{\end{equation*}}
\newcommand{\beqa}{\begin{eqnarray}}
\newcommand{\eeqa}{\end{eqnarray}}

\begin{document}
\title{Aiding Autonomous Vehicles with Fault-tolerant V2V Communication}

\author{Vladimir~Savic, Elad~M.~Schiller, and Marina~Papatriantafilou
\thanks{Copyright (c) 2017 IEEE. Personal use of this material is permitted. However, permission to use this material for any other purposes must be obtained from the IEEE by sending a request to pubs-permissions@ieee.org. The original version of this paper is submitted to IEEE journal for possible publication. This version is preliminary and subject to changes.}
\thanks{Authors are with the Dept. of Computer Science and Engineering (CSE), Chalmers University of Technology, Sweden. Emails: \{savicv, elad, ptrianta\}@chalmers.se}%
\thanks{This work was supported by Chalmers' Area of Advance Transport, as well as Vinnova's projects intelligent TRAffic maNagement System based on ITS (iTRANSIT), C-ITS Testplattform f\"{o}r framtidens transportsystem (CHRONOS steg1), and by the Swedish Foundation for Strategic research (framework project FiC). Parts of this work were presented at the IEEE Intl. Symp. on Intelligent Vehicles (IV'2017) \cite{Savic2017}.}%
}

\maketitle

\begin{abstract}
Vehicle-to-vehicle (V2V) communication is a key component of the future autonomous driving systems. V2V can provide an improved awareness of the surrounding environment, and the knowledge about the future actions of nearby vehicles. However, V2V communication is subject to different kind of failures and delays, so a distributed fault-tolerant approach is required for safe and efficient transportation. This work considers fully autonomous vehicles that operates using local sensory information, and aided with fault-tolerant V2V communication. The sensors provide all basic functionality, but are overridden by V2V whenever is possible to increase the efficiency. As an example scenario, we consider intersection crossing (IC) with autonomous vehicles that cooperate via V2V communication, and propose a fully distributed and a fault-tolerant algorithm for this problem. According to our numerical results, based on a real data set, we show the crossing delay is only slightly increased in the presence of a burst of V2V failures, and that V2V can be successfully used in most scenarios.
\end{abstract}

\section{Introduction}\label{sec:intro}

\textit{Autonomous vehicles} are expected to enable safer, greener, more efficient and more comfortable transportation \cite{Wymeersch2015,Hult2016}. They are equipped with a wide range of sensors \cite{Mukhtar2015}, such as Global Positioning System (GPS) receivers, radars, lidars, cameras, and inertial measurement unit (IMU). In addition, they should be equipped with radios for V2V wireless communication \cite{Hafner2013, casimiro2013} that would be used to exchange all relevant information with nearby vehicles and the infrastructure. V2V would facilitate increased awareness of surrounding environment, including the distant objects out of the sensing horizon. Moreover, the vehicles would be able to optimize their trajectory using the sensory information and the future trajectories from nearby vehicles. This work considers fully autonomous vehicles that operates using local sensory information and aided with fault-tolerant V2V communication. The sensors are expected to provide all basic functionality, while V2V would be used only when it is possible to increase the efficiency. 

Each year there are about 50 millions injures and deaths worldwide caused by road traffic crashes \cite{who2015}. A substantial part \cite{Azimi2013} happen on road intersections due to the vehicle collisions. Moreover, since traditional intersections are managed by traffic lights and stops signs, there would be an excessive delay with autonomous vehicles. On the other hand, since wireless communication is prone to failures and delays, a centralized intersection manager is not a desirable solution. We rather consider a \textit{distributed} approach in which the vehicles need to agree, using sensors and V2V communication, on the order in which they should cross the intersection. Although state-of-the-art provide solutions for many different problems (see Section \ref{sec:rel}), to the best of our knowledge, there is no solution that can handle an unknown and unlimited number of communication failures and delays. In contrast to these solutions, our hybrid proposal, based on both sensors and V2V, can handle an unknown number of communication failures, and can satisfy boththe safety and the liveness requirements. Our numerical results, based on a real data set, show that the crossing delay is only slightly increased in presence of a burst of V2V failures, and that the V2V can be successfully used (i.e. not overridden by sensors) in most scenarios.

The remainder of this paper is organized as follows. In Section \ref{sec:rel}, we provide the related work for the problem at hand, and in Section \ref{sec:smodel}, we formulate the problem and provide the required models. Then, in Section \ref{sec:distalg}, we provide our novel fault-tolerant and distributed algorithm for IC, and in Section \ref{sec:distalg}, we provide experimental results. Finally, in Section \ref{sec:conc}, we provide conclusions, and few suggestions for future work.

\section{Related Work}\label{sec:rel} 

We overview here the main methods on collision avoidance and intersection crossing for autonomous and semi-autonomous vehicles.

In \cite{Mukhtar2015}, the authors provide a survey on vehicle detection techniques, with a focus on computer vision. The sensors are classified into two groups: active (such as lasers, radars and lidars) and passive (such as cameras, and acoustic sensors), and then compared to each other in terms of range, cost and many other features. The radar is considered as the best active sensor, since it provides long-range ($>150 m$) real-time detection even under very bad weather (e.g., foggy, rainy) conditions. On the other hand, a radar is not able to determine the shape of the object, which can be done with lidar, a costly alternative. These problems encouraged authors to focus on passive sensors, such as cameras. Cameras are low-cost sensors, able to provide a very precise information about the objects. However, their main drawback is a high complexity of data processing, low range during nights, and sensitivity to weather conditions. Note that authors did not consider any kind of communication between vehicles that would resolve some of the sensors' problems.

In \cite{Hafner2013}, the authors use V2V for decentralized and cooperative collision avoidance for semi-autonomous vehicles, in which the control is taken from the driver once the car enters a critical area. The algorithm is tested using vehicles equipped with: differential GPS (DGPS), inertial measurement unit (IMU), dedicated short-range communication (DSRC) unit, and an interface with actuators. Their solution aims to compute an optimal throttle/brake control to avoid entering the capture area, in which no control action can prevent a collision. The estimation of longitudinal displacement, velocity and acceleration is performed using Kalman filtering. This estimation takes into account a bounded communication delay found experimentally (based on their experimental results, the worst case delay is 0.4s). Their experimental results showed that all collisions can be  avoided, and that the algorithm does not introduce a significant delay.

The work in \cite{Azimi2013} develops a reliable and efficient intersection crossing protocols using V2V communication. The proposed solutions are able to avoid deadlocks and collisions at intersections. The protocols are fully distributed since they do not rely on any centralized unit such as intersection manager. The autonomous vehicles are equipped with a similar set of sensors as in \cite{Hafner2013}, and also a DSRC unit for V2V communication. The vehicles interact with each other using standardized basic safety messages (BSM) adapted for intersection crossing problem. The communication failures are not explicitly handled since it is assumed that local sensors would be able to handle this problem. The proposed protocols are tested using AutoSim simulator/emulator, which utilizes a real city topography, and consists of control, communication and mobility modules. The results showed that the proposed protocols outperform the traditional traffic light protocols in terms of trip delay, especially when the traffic volume is asymmetric. Finally, this work is extended in \cite{Azimi2014} to account for GPS position inaccuracies, and also deal with roundabout intersections.

Cooperative collision avoidance with imperfect vehicle-to-infrastructure (and vice-versa) communication is analyzed in \cite{Colombo2015}. The centralized supervisor, located at the intersection, acquires the positions, velocities, and accelerations of the incoming stream of vehicles, and then decides either to allow vehicles' desired inputs, or to override them with a safe set of inputs. The communication is subject to failures, with the success reception probability based on the Rayleigh fading channel model. According to simulation results, the mean time between the accidents is significantly increased, but a collision may still happen if the override message has been lost.

In \cite{Dresner2008}, authors propose a novel centralized intersection crossing method (reffered to as AIM) for both autonomous and semi-autonomous vehicles. The vehicles and the intersections are considered as autonomous agents, which communicate via V2I/I2V communication links. The intersection agent uses its internal reservation policy to grant, reject or request modification of the vehicles' requests. Since this policy ensures that there will never be more than one vehicle in the conflict area, the collisions are not possible. However, in case of unbounded communication failures, an indefinite delay may happen. According to their results, their approach is much faster than traditional methods based on traffic lights and stop signs.

A hybrid centralized/distributed architecture that ensures both the safety (no collisions), and the liveness (a finite crossing time), is proposed in \cite{Kowshik2011}. They assumed that there are no stop signs and traffic lights at the intersection, and that the vehicles are equipped with a positioning unit, internal sensors, and a V2V communication unit. To resolve the problem with a bounded communication delay and packet losses, the rear car needs to break with maximum deceleration. They compared the proposed solution with stop-sign and traffic-light technologies and found that the average travel time is significantly reduced.

In \cite{Steinmetz2014}, the authors consider intersection crossing using the controller located in the intersection center. The analysis is provided for both uplink and downlink of the imperfect communication channel used to relay the positions, velocities and the destination of all approaching vehicles. The communication is carried out via frequency division duplexing, and multipath fading is modeled with Rayleigh distribution. Then, by minimizing the failure probability, they found the optimal values of the transmit power, the number of channels, and the communication rate.

The work in \cite{Colombo2012} focuses on semi-autonomous vehicles with the controller that can override the driver and prevent potential collision at the intersections. They define this problem as a scheduling problem and then solve it by determining the largest set of states for which there exists a control that guarantees collision avoidance (known as maximal controlled invariant set). The original problem is NP-complete, but the proposed approximation has a polynomial complexity.

Finally, in \cite{Makarem2012}, the main contribution is a novel decentralized navigation function for autonomous vehicles with a predefined path. It takes into account the expected arrival time to the intersection, which results in a fluent traffic without unnecessary stops. The weighting factors are introduced to give a higher priority to heavier vehicles. The proposed decentralized solution is compared with a centralized one and traffic lights, in terms of energy consumption and the maximum throughput. The key result is a higher throughput, but the price is higher energy consumption than a centralized solution.

Since this is not an exhaustive list of contributions in this area, we also refer the readers to a recent survey  \cite{Chen2016a} on cooperative intersection management.

\section{System Model}\label{sec:smodel}

\subsection{Preliminaries}\label{subsec:prel}
\begin{figure}[!t]
\centerline{
\includegraphics[width=0.6\textwidth]{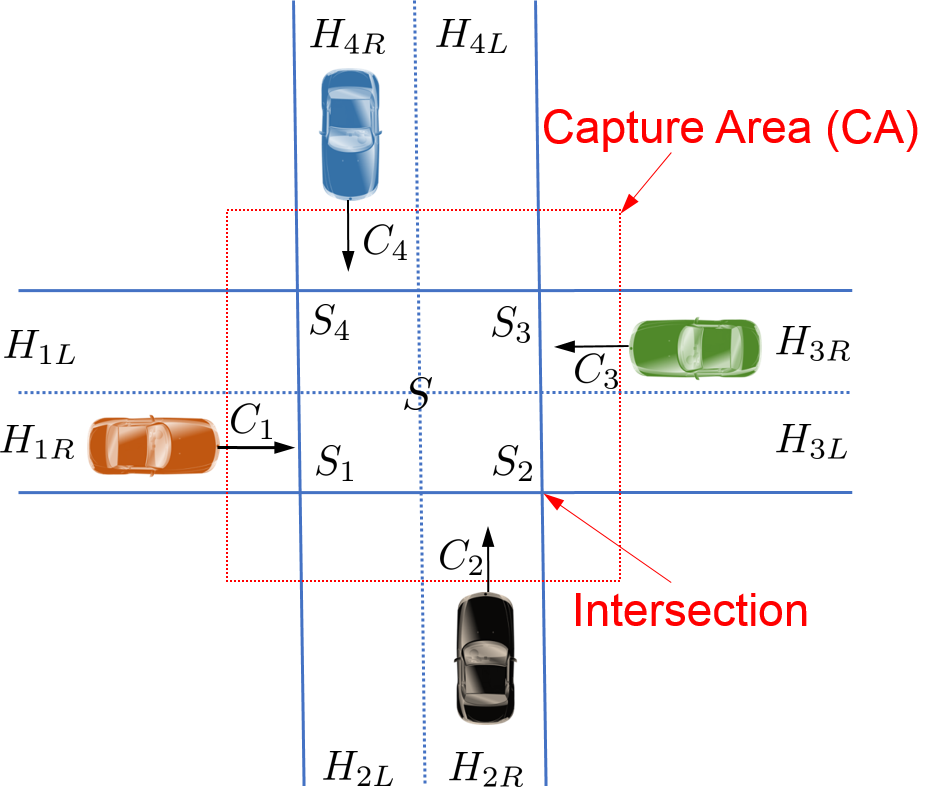}}
\caption{An illustration of 2-lane road intersection with four incoming cars.}
\label{fig:intersec4cars}
\end{figure}

We consider $N_C$ fully autonomous cars or platoons ($C_1, \ldots, C_{N_C}$) on different lanes, competing to cross a road intersection as shown in Fig. \ref{fig:intersec4cars}. They are equipped with all required sensors (such as cameras, radars, GPS receiver, etc.) that can provide a basic functionality \cite{Mukhtar2015,Skog2009,Vu2013}. More precisely, they are able to:
\bi
\item estimate its own position, velocity and acceleration,
\item detect all surrounding cars, pedestrians, and obstacles,
\item detect the lanes and the intersections,
\item perform automatic cruise control, platooning, and intersection crossing (IC),
\item measure the distance and the angle to other vehicles (i.e., their positions).
\ei

We refer to this system as a \textit{sensor detector (SD)} system. Moreover, the cars have an equal size and weight, and a unique identifier. Initially, all the cars are moving towards the intersection and no car has a priority to cross the intersection. The road has 2 lanes, so the intersection ($S$) can be divided into four subsections ($S_1, S_2, S_3, S_4$).\footnote{These assumptions are made to facilitate the presentation of our problem, but an extension to more complex models is not difficult.} In each of these subsections, collision may occur if two cars occupy it simultaneously. Since cars' acceleration is limited by their inertia, we also define a \textit{capture area (CA)}, i.e., the area in which no control action can stop the entrance to the intersection. Note that the capture area is not constant, and it depends on the cars' current dynamics.

We then define the following variables:
\bi
\item[] $UID_j$ - unique identifier of $C_j$,
\item[] $x_{j}^t$ - true longitudinal (1D) position of $C_j$ at time $t$,
\item[] $v^t_{j}$ - true longitudinal (1D) velocity of $C_j$ at time $t$,
\item[] $a^t_{j}$ - true longitudinal (1D) acceleration of $C_j$ at time $t$,
\item[] $x_{S}$ - the central point of the intersection $S$,
\item[] $CLANE_j$ - current lane, before crossing the intersection ($CLANE_j
\in \{H_{1R}, H_{2R}, H_{3R}, H_{4R}\}$),
\item[] $NLANE_j$ - next lane, after crossing the intersection ($NLANE_j
\in \{H_{1L}, H_{2L}, H_{3L}, H_{4L}\} \backslash H_{jL}$).
\ei
The index $t=1,\ldots,N_t$ represents the discrete time slot, and the time interval between two time slots is denoted with $T$. Both sensing and V2V communication units use the same time slot. Note that the time indexes are omitted for variables that remain constant with time.

\subsection{Messages and failures}\label{subsec:msgs}

Autonomous cars with a SD, described in the previous section, is able to ensure safe\footnote{Although the safety is not guaranteed, they are much safer than manual-driving cars.} IC, but in an inefficient way since sensors typically have low sensing horizon (100-200 m), sometimes require a complex data processing, and do not have access to the future trajectories of the detected cars. Therefore, we would like to override them with V2V whenever possible. So we assume that each car has a radio transceiver available, and can periodically transmit and receive the messages within a communication range $R$.

We use two different messages for this problem: the ENTER message, and the ACK message (denoted by $MSGENTER$ and $MSGACK$, respectively). These messages should include all relevant information required for safe and efficient IC. We use here a similar set of messages as in \cite{Azimi2013}, which are defined according to DSRC SAE J2735 standard \cite{sae2009}. However, to adapt to our problem, we do not transmit the data not needed for IC (such as trajectory list), but this data should be sent for other tasks of autonomous vehicles.

The format of the messages for car $j$ is given as follows:
\bn
\item ENTER message: \\$~MSGENTER^t_j=$ \\$~\{UID_j,MSGTYPE_j,CLANE_j,NLANE_j,\tau^t_{{\rm{MTI}},j}\}$
\item ACK message: \\$~MSGACK^t_j= \{UID_j,MSGTYPE_j\}$
\en
where $MSGTYPE_j \in \{{\rm{'ENTER'}},{\rm{'ACK'}}\}$, and the parameter $\tau^t_{{\rm{MTI}},j}$ serves for priority management and will be defined later. Note that, in contrast to \cite{Savic2017,Azimi2013}, cross and exit messages are not needed since these tasks will be handled by sensor detectors.

Once car $C_j$  gets close enough to the capture area, it sends the $MSGENTER$. Since this is a safety-critical problem, this message will be sent at least as soon as the following condition is satisfied:
\be\label{eq:sendEnter}
Prob~\{x^{t+t_{j}}_j\in CA\} \ge \epsilon
\ee
where $\epsilon$ is the desired tolerance (e.g., $\epsilon=10^{-9}$), and $t_{j}$ is the number of time slots before $C_j$ gets the intersection. This probability needs to be estimated at each time slot before $C_j$ sends the $MSGENTER$, and can be easily computed from the predictive probability distribution, which can be found via Kalman or Particle filtering \cite{Arulampalam2002}. %

The variable $t_{j}$ should be set to the value that would allow car to start communication with other cars as soon as it is within the communication range ($R$) of CA. For example, given the current velocity ($v^t_{j}$), and assuming zero acceleration ($a^t_{j}=0$), we can set $t_{j}=\lceil R/(v^t_{j}\cdot T)\rceil$ where $\lceil~\rceil$ is the ceiling operator.

Regarding $MSGACK$, it will be sent to acknowledge that all $MSGENTER$s from other competing cars are received. Once all $MSGACK$s are received, the car is allowed to make a decision about the priority, as described in more details in the next sections.

Finally, we also assume the following: 
\bi
\item Cars can experience an \textit{unknown} number of consecutive receive-omission failures (i.e., fail to receive the message). Without loss of generality, we consider one burst of errors, and denote it by $f_j$ ($f_j \ge 0$) for car $C_j$.
\item If the number of failures is \textit{larger than} a predefined threshold $F$, the car will use its sensor detectors instead of V2V. 
\item Cars are able to \textit{successfully} transmit all messages, and the delivered packet does not contain erroneous data.
\item Cars are \textit{fully cooperative} and they never send malicious messages. 
\item Each message ($MSGENTER$ and $MSGACK$) is sent \textit{within one} packet, so it will be either fully delivered, or completely lost.
\item If the message is not received during \textit{the same} time slot in which is sent, it is considered outdated and \textit{discarded}.
\ei

Therefore, these assumptions allow us to focus on the most frequent failures caused by obstructed wireless channel (e.g., non-line-of-sight, jammers, interference). We also do not make any assumption about the channel model (such as Rayleigh fading \cite{Colombo2015}), nor predefine the number of failures. However, we do not consider send-omissions, nor erroneous data, since these problems can be significantly mitigated by testing the transmitters, and using an appropriate error-correcting code \cite{Kowshik2011}. Other problems, such as non-cooperative or malicious behavior, are out of focus of this paper, and can be partially resolved using another type of algorithms \cite{Raynal2010}.

\subsection{Control actions}\label{subsec:ctrl}

We first define the parameter $\tau^t_{{\rm{MTI}},j}$ that will be used to determine the priority for our IC problem. One may assign the priorities a priori based on the type and the importance of the car (e.g., a police or an emergency car would go first), but this would cause an additional delay. Although there are many alternatives \cite{Tonguz2016, Azimi2013}, we use \textit{first-come-first-served (FCFS)} approach based on the current position estimate and the cars' dynamics. Thus, we use the \textit{mean time to intersection (MTI)} of $C_j$ at time $t$, that can be found as:
\be\label{eq:mti}
\tau^t_{{\rm{MTI}},j}=\frac{-v^t_{j}+\sqrt{(v^t_{j})^2+2a^t_{j}({x}_{S}-\hat{x}^{t}_j)}}{a^t_{j}}
\ee
The car with lowest MTI will first cross the intersection, while the other cars would need to wait for the exit of all cars with lower MTIs. In the rare situation, in which MTIs are equal, we use instead \textit{UIDs} as a tie-breaker. Note that the priority management is not a safe-critical operation, so we can use the expected value instead of the worst-case estimate. 

We then define here the control action that should be performed after the agreement is established (denoted by $MAINCTRL$). Note that a feasible control action exists \cite{Hafner2013}, since we ensured that the communication started before entering the CA. Moreover, since there are many possible control actions, we assume that cars slow-down/speed-up with a constant acceleration/deceleration. We neglect the velocity and acceleration errors, but not the position errors since they may be large.

\begin{figure}[!tb]
\centerline{
\includegraphics[width=0.6\textwidth]{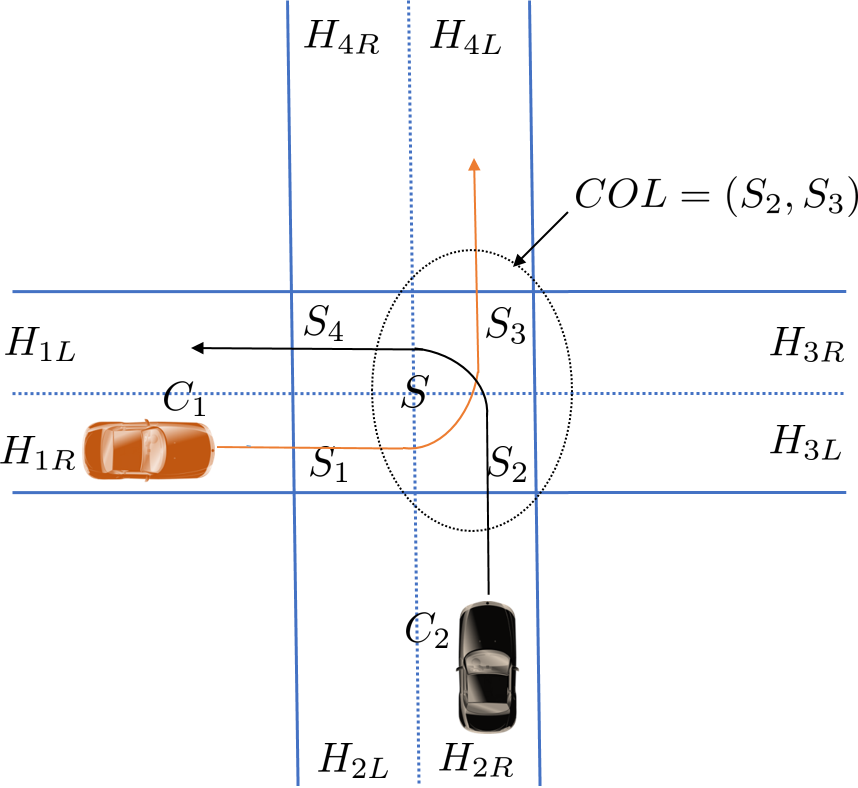}}
\caption{An illustration of collision area for a given route of the cars.}
\label{fig:colArea}
\end{figure}

Once the IC algorithm is performed, all competing cars have access to others' cars $MSGENTER$s. Therefore, using $\tau^t_{{\rm{MTI}},j}$, $CLANE_j$ and $NLANE_j$ ($j=1,\ldots,N_C$), cars can determine if collision is possible. There are two situations in which the collision cannot happen: (i) cars never occupy the same subsection, and (ii) cars do not occupy the same subsection simultaneously. Otherwise, the collision is likely to happen and only one car is allowed to cross. The collision area ($COL$) depends on the cars' routes and may be any of the subsections $S_i$ ($i=1,\ldots,4$) or a combination of them. An example is shown in Fig. \ref{fig:colArea}. It is also possible that the collision area is empty ($COL = \emptyset$), e.g. if all cars turn right. This is exactly a situation in which V2V provides a faster crossing than SD.

Therefore, $MAINCTRL$ should allow $C_j$ to proceed with its desired acceleration either if $COL = \emptyset$ or $\left|\tau^t_{{\rm{MTI}},j}-\min_{i \ne j}{(\tau^t_{{\rm{MTI}},i})}\right|>\tau_{TH}$ where $\tau_{TH}$ is the threshold that depends on cars' dynamics, and the minimum safety distance. Otherwise, the collision is possible, so $C_j$ can proceed only if $\tau^t_{{\rm{MTI}},j}<\min_{i \ne j}{(\tau^t_{{\rm{MTI}},i})}$ or $\tau^t_{{\rm{MTI}},j}=\min_{i \ne j}{(\tau^t_{{\rm{MTI}},i})}$ \& $UID_j>\min_{i \ne j}{(UID_i)}$. As we can see, we used FCFS and the identifiers as a tie-breaker. Once a car cross the intersection, it will be removed from the list, and the remaining cars would repeat the procedure.

Whether $C_j$ has a priority or a collision is not possible, it can keep moving (until cross the intersection) with the desired acceleration $a_{j,PR}$, for example, equal to the current acceleration:
\be\label{eq:pr}
a_{j,PR}=a_j^{t+1}=a_j^{t+2}=\ldots = a_j^{t}
\ee
Otherwise, it needs to slow down, but just as little as necessary to avoid collision. Assuming that we want to reduce cars' displacement for $D$ (which should be at least equal to the width of the $COL$), this acceleration ($a_{j,NOPR}$) can be found using standard kinematic equations:
\be\label{eq:nopr}
a_{j,NOPR}=a_j^{t+1}=a_j^{t+2} = \ldots = a_j^{t}-\frac{2D}{(\tau^{t+1}_{j,COL})^2}
\ee
$\tau^{t+1}_{j,COL}$ is the worst-case remaining time to the collision area, and is given by:
\be\label{eq:timeCol}
\tau^{t+1}_{j,COL}=\frac{-v^{t+1}_{j}+\sqrt{(v^{t+1}_{j})^2+2a_{j,PR}({x}_{COL}-\hat{x}^{t+1}_{j,MAX})}}{a_{j,PR}}
\ee
where ${x}_{COL}$ is the entrance point of the collision area, and $\hat{x}^{t+1}_{j,MAX}$ is the worst-case position estimate of $C_j$. Note that this computation needs to be done at time $t$, so it is assumed that the car does not change the acceleration (i.e. it is equal to $a_{j,PR}$, what would lead to a collision). Once $C_j$ gets the priority, it will increase the acceleration to the desired value, e.g. to the same value ($a_{j,PR}$) as in \eqref{eq:pr}.

Finally, in case of too many failures (larger than $F$), SD would be used instead of V2V, and we do not provide here the detailed priority management for that case. In general, the decisions would be based on all useful sensed information, such as positions of all surrounding cars.

\section{Distributed fault-tolerant IC algorithm}\label{sec:distalg} 
Given the models and the assumptions from Section \ref{sec:smodel}, we now propose a distributed fault-tolerant algorithm that ensures safe and efficient IC in the presence of unknown number of communication failures. The algorithm consists of three parts: i) main SD-based algorithm before entering the IC, ii) V2V-based ENTER algorithm, and iii) SD-based EXIT algorithms.

\subsection{Overriding sensor detector (SD)}\label{subsec:override}

\begin{figure}[!t]
\centerline{
\includegraphics[width=0.6\textwidth]{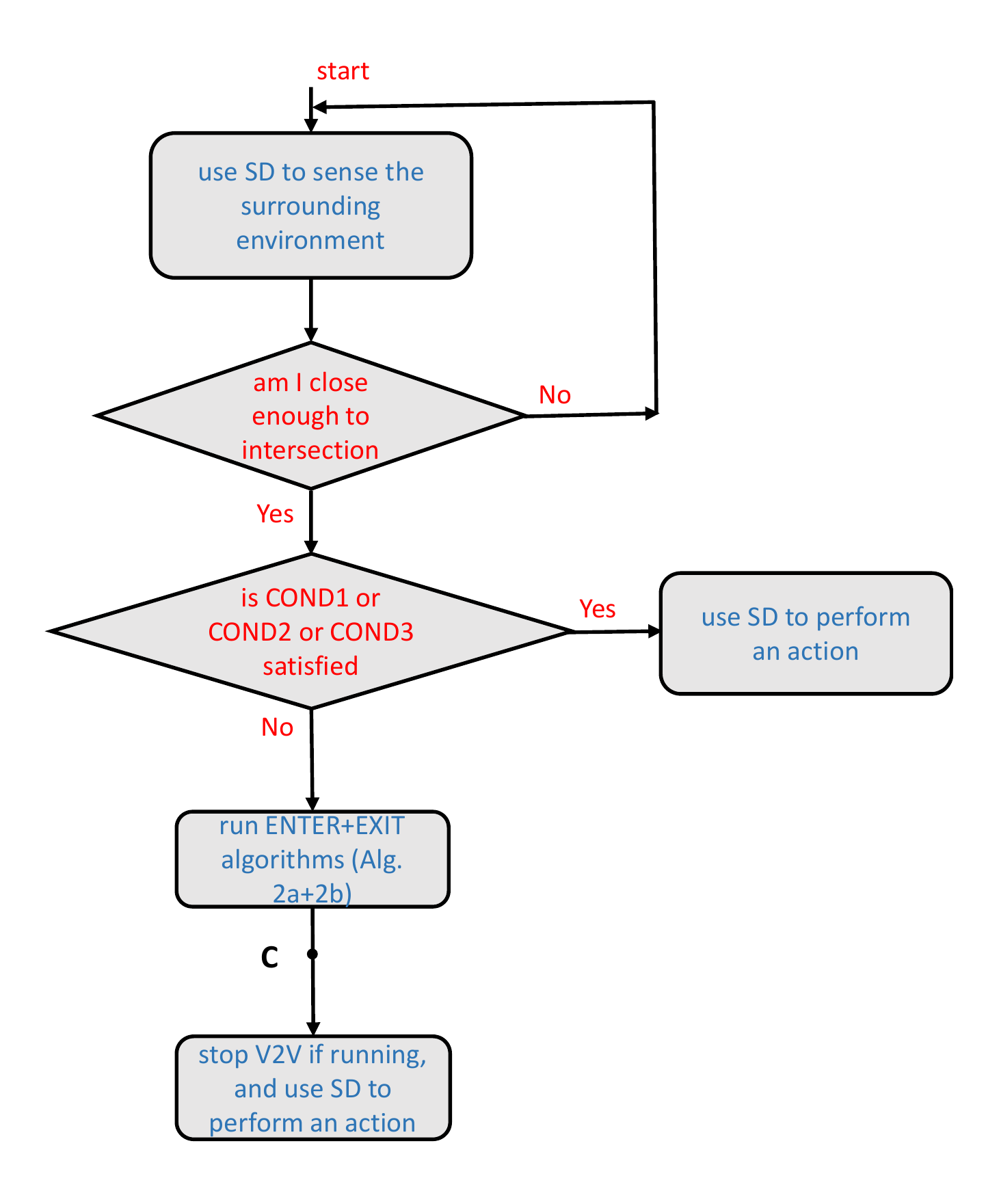}}
\caption{Alg. 1: Flowchart of the main SD-based algorithm for car $C_j$. The time and car indexes are omitted for ease of presentation.}
\label{fig:alg1-sd}
\end{figure}

A car approaching intersection uses its SD to: estimate its own position and the dynamic, to detect all objects in front of the car and around the intersection, and to estimate their positions. Once close enough to intersection, it uses that information to make a decision if it is possible to switch to V2V. We define the following conditions from the point of view of one car:
\bi
\item COND1: no vehicles/objects within the predefined radius from intersection on all incoming lanes,
\item COND2: there is at least one vehicle, object or pedestrian on the same lane in front of me,
\item COND3: there is at least one vehicle on other lanes (excluding mine) already competing to cross the intersection. 
\ei

Even if one of these conditions is satisfied, the car would keep using SD to perform an action. For example, it there are no vehicles nor other obstacles around the intersection (COND1), it would just carefully cross it. Otherwise, if there is an obstacle on the same lane (COND2), it would slow down to avoid the collision. If that obstacle is an autonomous vehicle, it would simply join the platoon and follow its decisions. Finally, if there are no obstacles on the same lane, but other autonomous vehicles already competing to cross the intersection (COND3) (those vehicles should signal it by turning on a specific light that can be detected by camera), the car would need to wait for at least one vehicle to exit the intersection.

If none of the conditions is satisfied, the car would switch to V2V mode. SD would be still turned on to avoid unforeseen situations and run other tasks (such as algorithm for pedestrian detection), but its IC algorithm would be overridden by V2V. This algorithm is summarized in Fig. \ref{fig:alg1-sd}.

\subsection{ENTER and EXIT algorithms}\label{subsec:override}

\begin{figure}[!tb]
\centerline{
\includegraphics[width=0.6\textwidth]{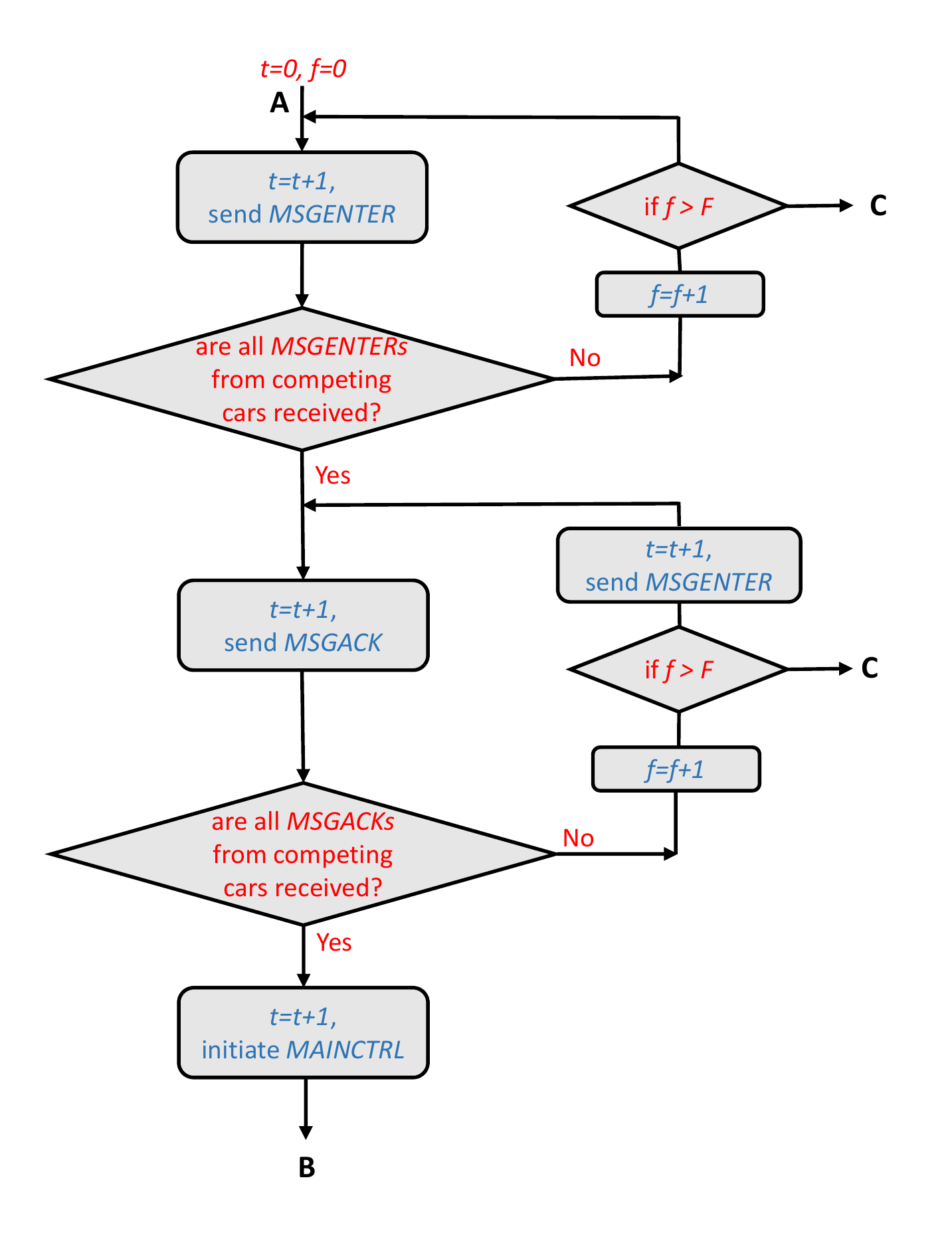}}
\caption{Alg. 2a: Flowchart of V2V-based ENTER algorithm for car $C_j$.}
\label{fig:alg2-enter}
\end{figure}

The ENTER algorithm for car $C_j$, summarized in Fig. \ref{fig:alg2-enter}, uses V2V to establish a consensus between all competing cars, and then decide whether to cross the intersection or wait for the next chance.

Once $C_j$ switched to V2V, it resets its time and failure counters ($t=0, f=0$) and attempt to send $MSGENTER$. Then, it checks if $MSGENTER$s from competing cars are received. If not, it increments the failure counter, and if it exceeds $F$, it switches back to SD mode. The messages may not be received either because of the failure, or because at least one of the competing car is still waiting to enter V2V mode. In that case, and if $f \le F$, $C_j$ repeats sending of $MSGENTER$. Once all $MSGENTER$s are received, $C_j$ will send $MSGACK$ to acknowledge that all required messages are received. If other $MSGACK$s are not received, $C_j$ will again increment the counter (the same one used for $MSGENTER$), and if $f \le F$, it will attempt again to send both $MSGENTER$ and $MSGACK$. This is done to avoid deadlock since $C_j$ cannot be sure which problem appears on other cars. Otherwise, $C_j$ is aware that all cars (including itself) have the same set of $MSGENTER$s. Therefore, using information from these messages, all cars will initiate $MAINCTRL$, which will ensure that only one car can cross the intersection. 

It is also important to point out that this algorithm will ensure that all competing cars either use V2V or SD, with an exception during a limited number of time slots. This is easy to show by assuming that $C_j$ commits $F+1$ failures and switch back to SD. From that point, it will stop transmitting V2V messages, so all other cars will assume that there is a problem, and after $F+1$ time slots also switch back to SD mode.\footnote{Note that our algorithm do not support switching again to V2V for the same intersection.} Therefore, the inconstancy will only appear for $2F+2$ time slots, so the collision can be avoided if $F$ is set to appropriate value that depends on the time slot duration ($T$) and the worst-case car's dynamics.

\begin{figure}[!tb]
\centerline{
\includegraphics[width=0.7\textwidth]{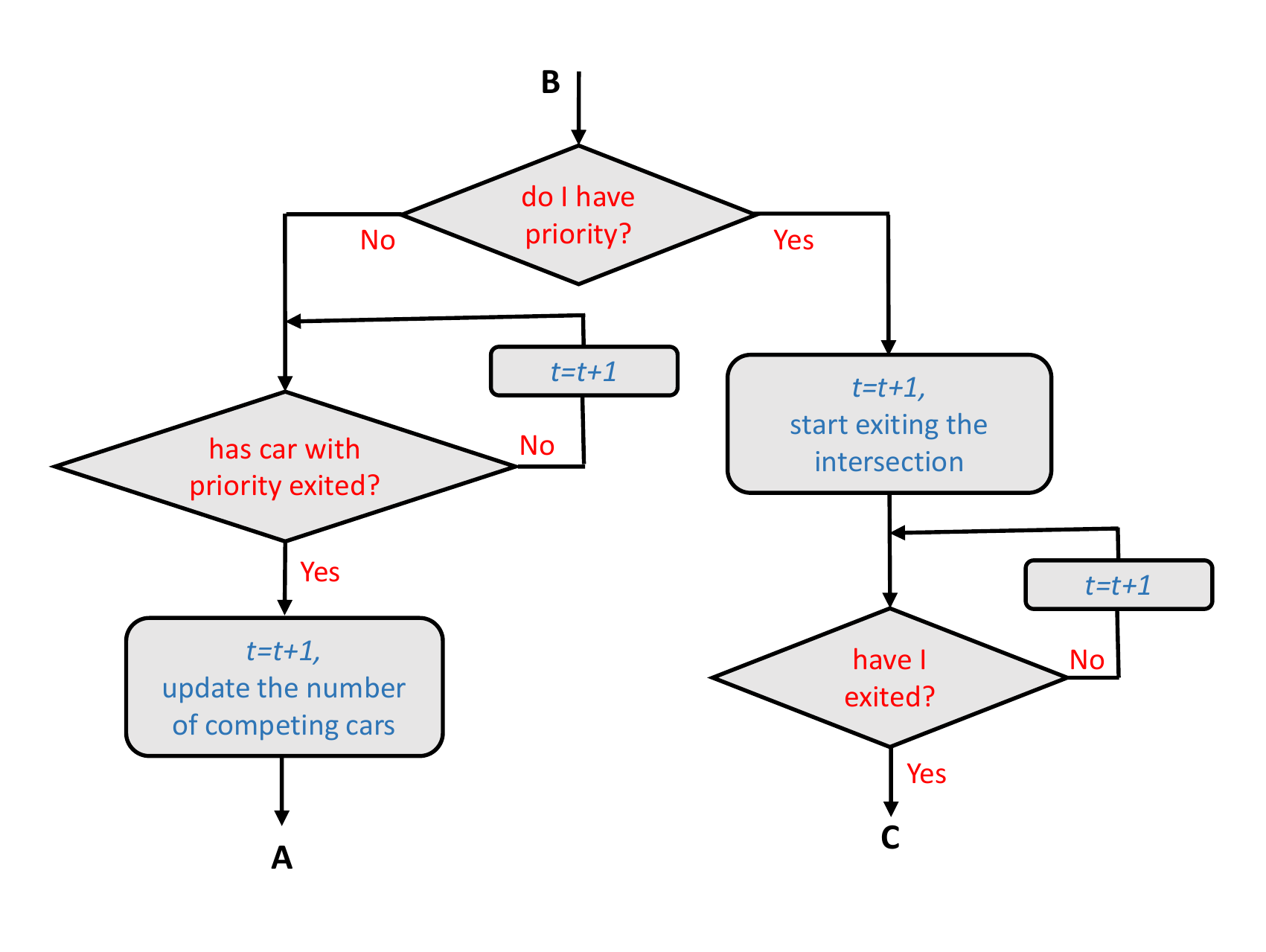}}
\caption{Alg. 2b: Flowchart of SD-based EXIT algorithm for car $C_j$.}
\label{fig:alg2-exit}
\end{figure}

Once $MAINCTRL$ is initiated, $C_j$ starts SD-based EXIT algorithm, which is summarized in Fig. \ref{fig:alg2-exit}. The car with priority will simply exit the intersection after some time (the number of time slots depends on cars' dynamics), and once it is aware of it (based on its position estimate), it will switch back to SD and continue with other tasks. The cars without priority will wait for the car with priority to exit, then update the number of competing cars using its SD, and finally start again the ENTER algorithm. Note that V2V is not used here since it would not increase the efficiency (i.e., no need for future trajectory nor long range sensing).

\subsection{Examples of time diagrams}\label{subsec:timeDiag}

\begin{figure}[!t]
\centerline{
\subfloat[]{\includegraphics[width=0.75\textwidth]{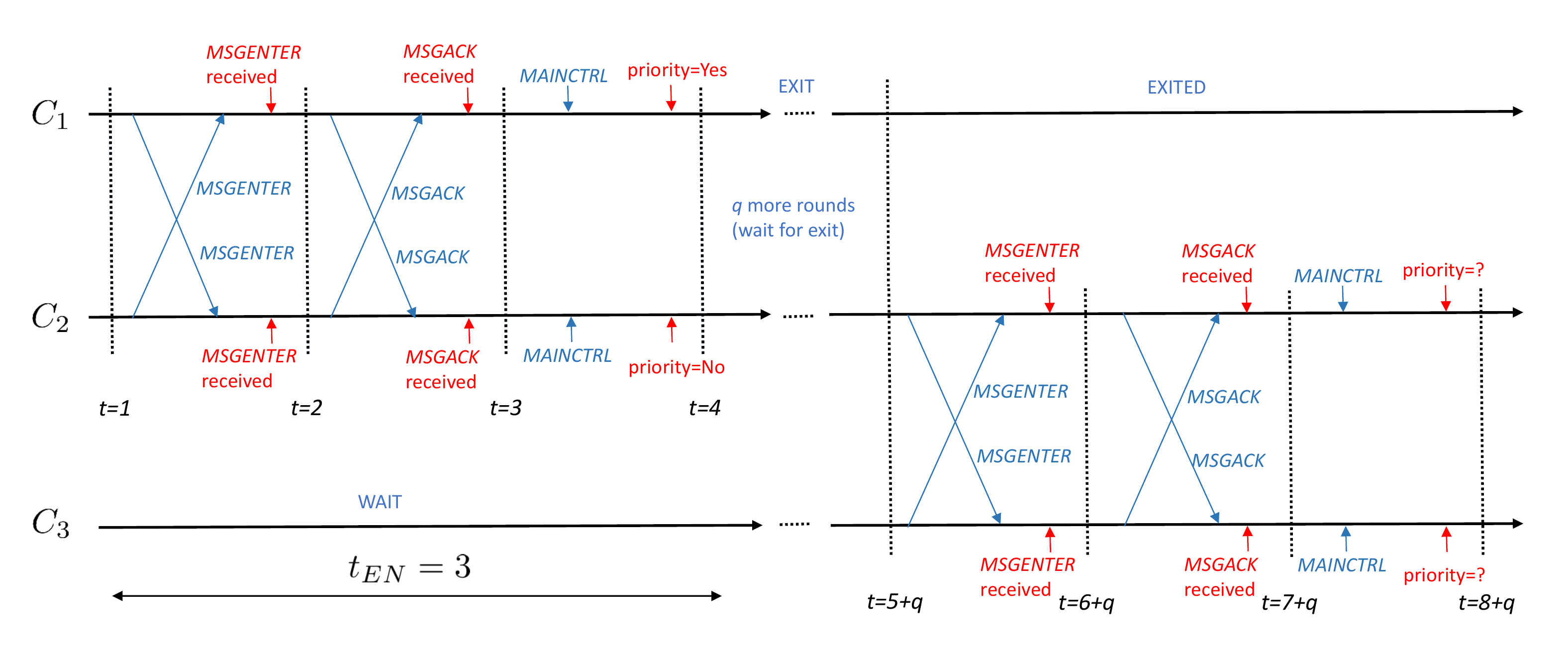}\label{fig:diagFull3cars}}
}
\centerline{
\subfloat[]{\includegraphics[width=0.59\textwidth]{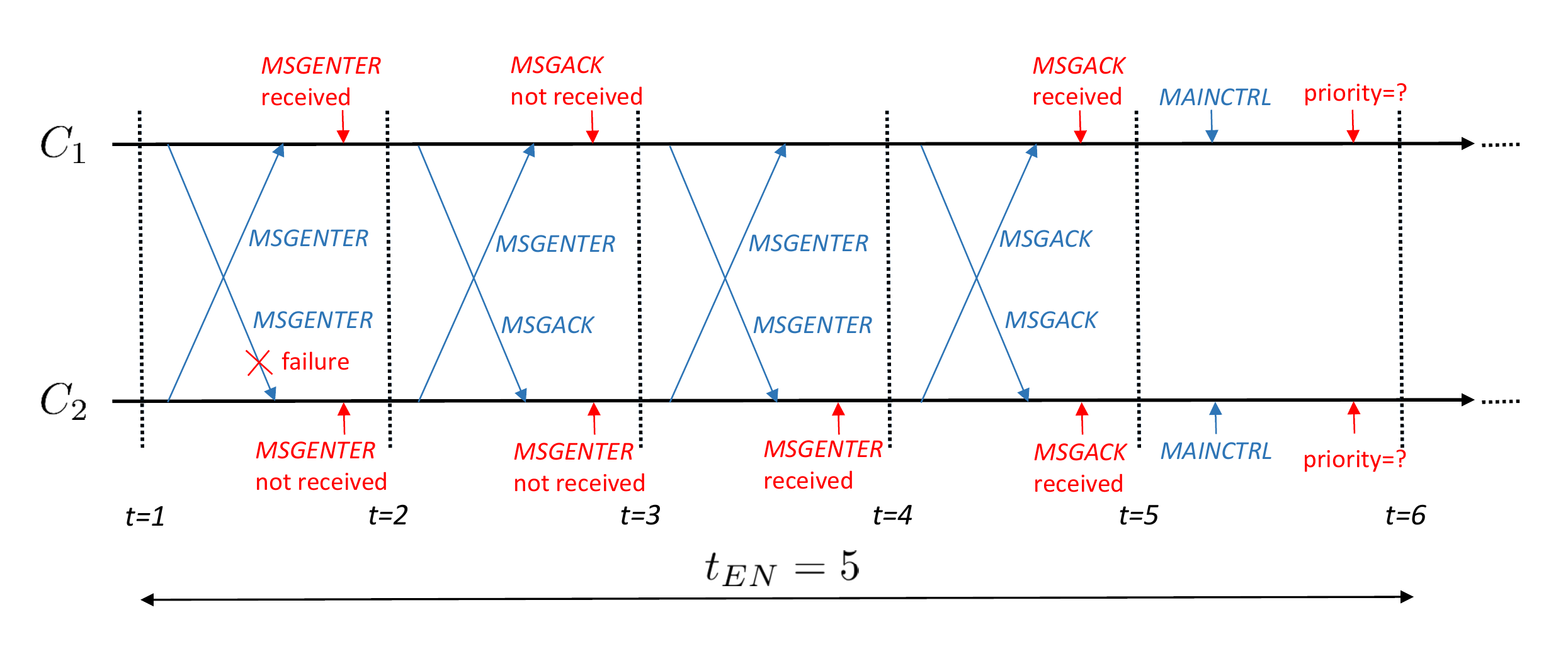}\label{fig:diag01Fail}}
}
\centerline{
\subfloat[]{\includegraphics[width=0.75\textwidth]{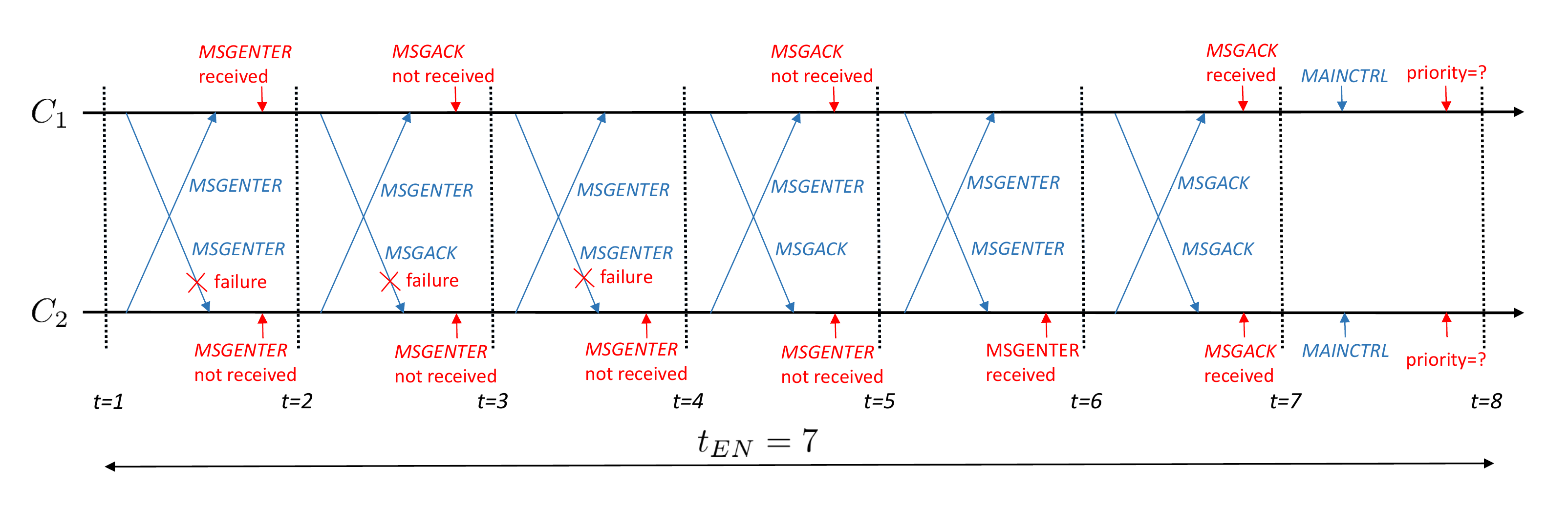}\label{fig:diag03Fail}}
}
\centerline{
\subfloat[]{\includegraphics[width=0.59\textwidth]{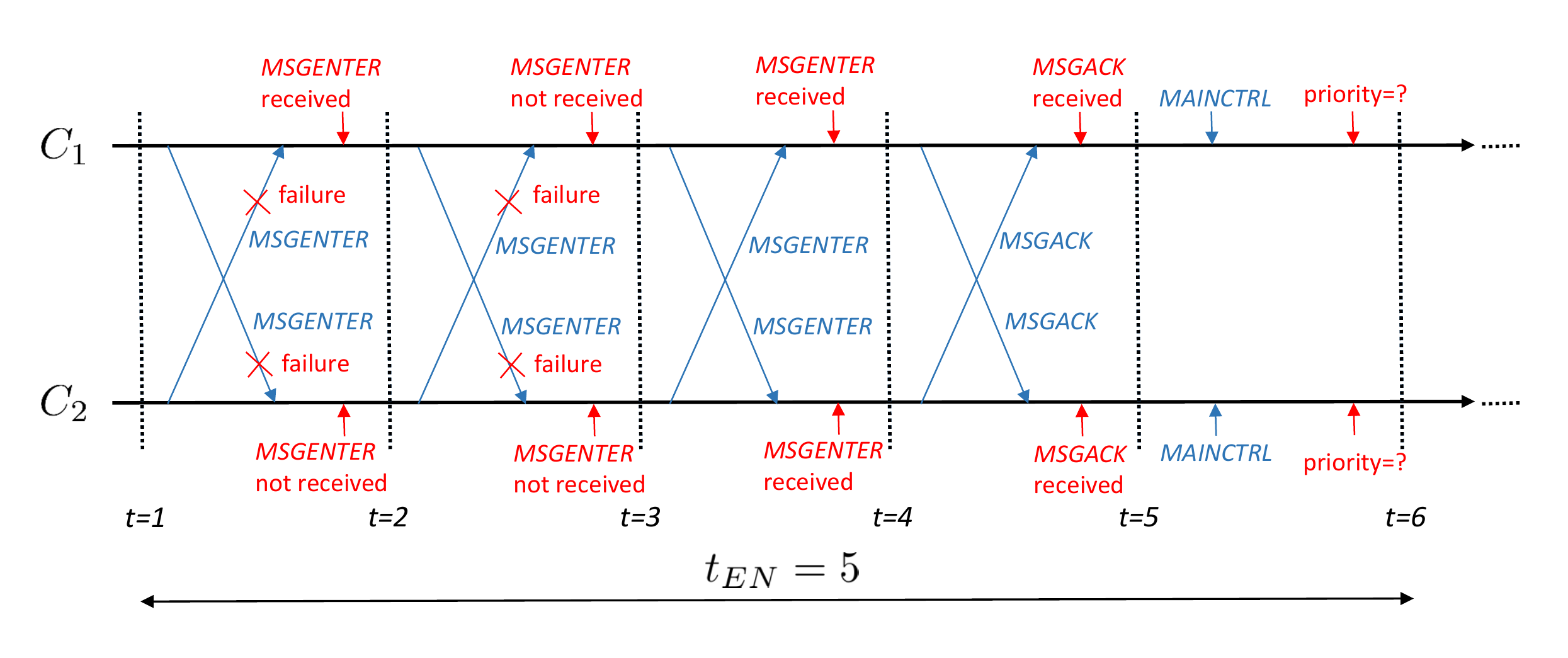}\label{fig:diag22Fail}}
}
\caption{Time diagrams of the ENTER algorithm for different number of failures: (a) 3 cars, no failures, (b) 2 cars, $(f_1, f_2)=(0,1)$, (c) 2 cars, $(f_1, f_2)=(0,3)$, and (d) 2 cars, $(f_1, f_2)=(2,2)$. }
\label{fig:diagramsFail}
\end{figure}

We analyze here time diagrams for ENTER algorithm which is the only part which uses V2V. We start with the example (Fig. \ref{fig:diagFull3cars}) for the execution without failures. We assume that there are three cars ($C_1, C_2, C_3$), but $C_3$ arrived to intersection little later. Therefore, $C_1$ and $C_2$ start to compete via V2V. They exchange $MSGENTER$s, acknowledge them via $MSGACK$ and finally initiate $MAINCTRL$. Therefore, the priority is determined in the same time slot ($t=4$), with the delay of only $t_{EN}=3$. Then, crossing the intersection will take $q$ time slots, which depends on the cars' dynamic (e.g. if $T=100$ ms, and the crossing time takes 3 s, the crossing would take $q=30$ time slots). After initiating $MAINCTRL$, $C_1$ would get priority (as an example), while $C_2$ would need to slow down little bit to avoid collision (assuming that $COL \ne \emptyset$) and start competing with $C_3$ using the same algorithm. Eventually, all 3 cars will cross the intersection.

We now move to the examples with communication failures. As shown in Fig. \ref{fig:diagramsFail}, we consider a scenario with 2 cars and the following number of failures: $(f_1, f_2)=(0,1)$, $(f_1, f_2)=(0,3)$ and $(f_1, f_2)=(2,2)$. 

In the first example (Fig. \ref{fig:diag01Fail}), $C_2$ fails to receive once $MSGENTER$, so it will attempt to send it again. Meanwhile, $C_1$ received the $MSGENTER$, so it can transmit $MSGACK$ to confirm it. $C_1$ also expects to receive $MSGACK$ from $C_2$, but it will receive $MSGENTER$ instead, and figure out that there is a failure at $C_2$. Consequently, $C_1$ will send again $MSGENTER$, which will be then received by $C_2$. Then, both cars can send $MSGACK$, which will be received by other car. Finally, both cars will initiate $MAINCTRL$ in the same time slot and decide about the priority. The total delay in this example is $t_{EN}=5$. 

In the second example (Fig. \ref{fig:diag03Fail}), $C_2$ fails to receive the messages for three consecutive time slots. We note that in the second round $MSGACK$ message is not received by $C_2$ in contrast to previous example. Since this message is not needed ($MSGENTER$ is expected), the second failure would not cause an extra delay.\footnote{This is the reason why we skipped the case with $(f_1, f_2)=(0,2)$ which has the same delay as $(f_1, f_2)=(0,1)$.} However, the third failure will cause extra two time slots since a new $MSGENTER$ will not be available in the time slot following this failure. Then, the last four time slots of the diagram are the same as in the previous example, and the total delay is $t_{EN}=7$. 

In the third example (Fig. \ref{fig:diag22Fail}), both cars fail to receive the message for two time slots. Therefore, both cars will keep sending $MSGENTER$, and will receive them in the third time slot. Then, they will simultaneously acknowledge it and initiate $MAINCTRL$. We see that the total delay is the same as in the first example ($t_{EN}=5$), so it is not affected by simultaneous failures. 

In summary, the total delay depends only on the maximum number of failures, and only an odd failure (3, 5, 7, ...) increases the delay for two extra time slots. Therefore, it follows by induction that the total delay of the ENTER algorithm is given by: $t_{EN}=\min{(F,2\lceil(\max_{i}{f_i})/2\rceil)}+3$. Note also that the SD will introduce an additional delay, caused by sensor latency and the data processing.

\begin{figure}[!t]
\centerline{
\subfloat[]{\includegraphics[width=0.5\textwidth]{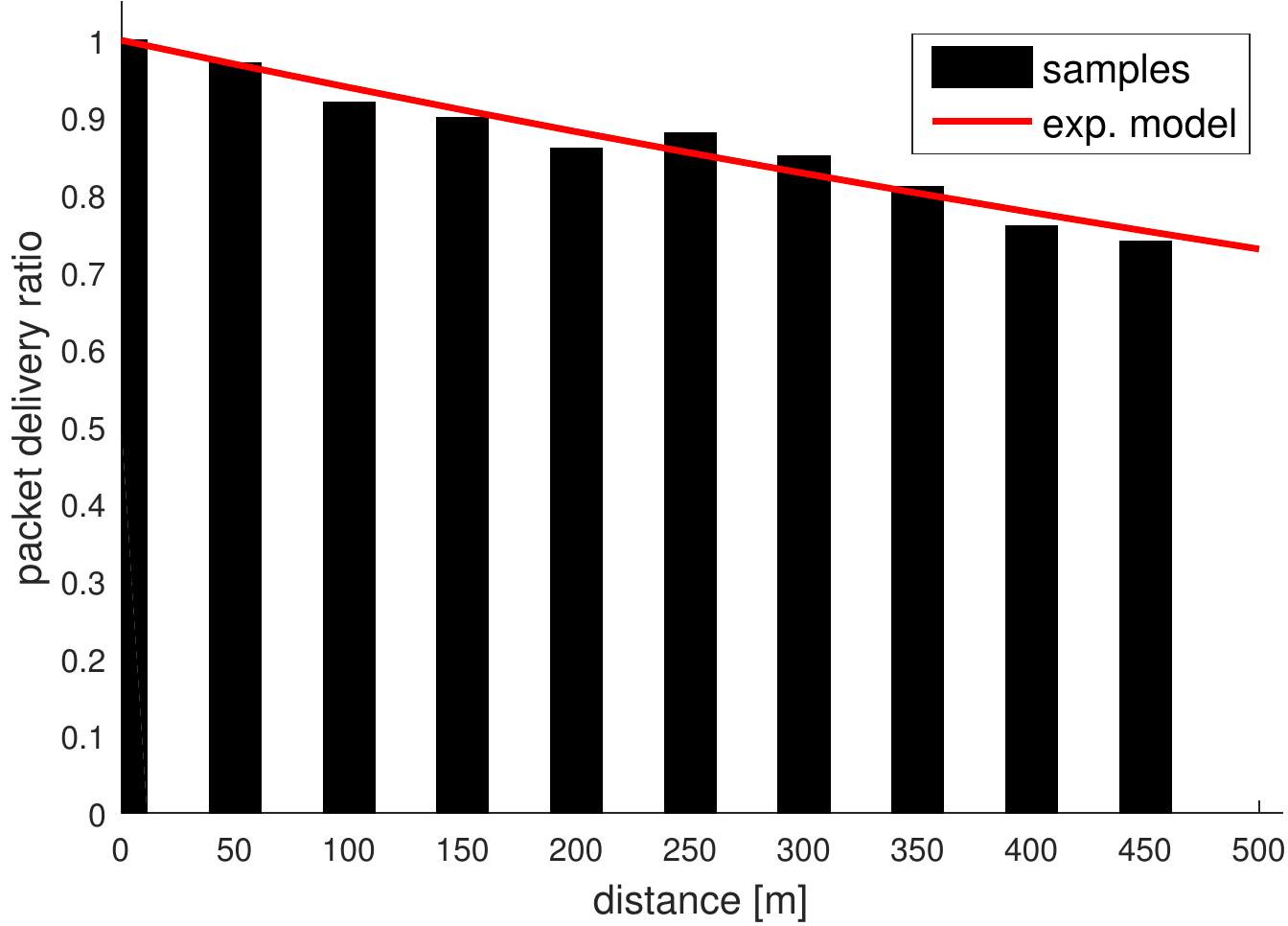}\label{fig:failDistOpen}}
}
\centerline{
\subfloat[]{\includegraphics[width=0.5\textwidth]{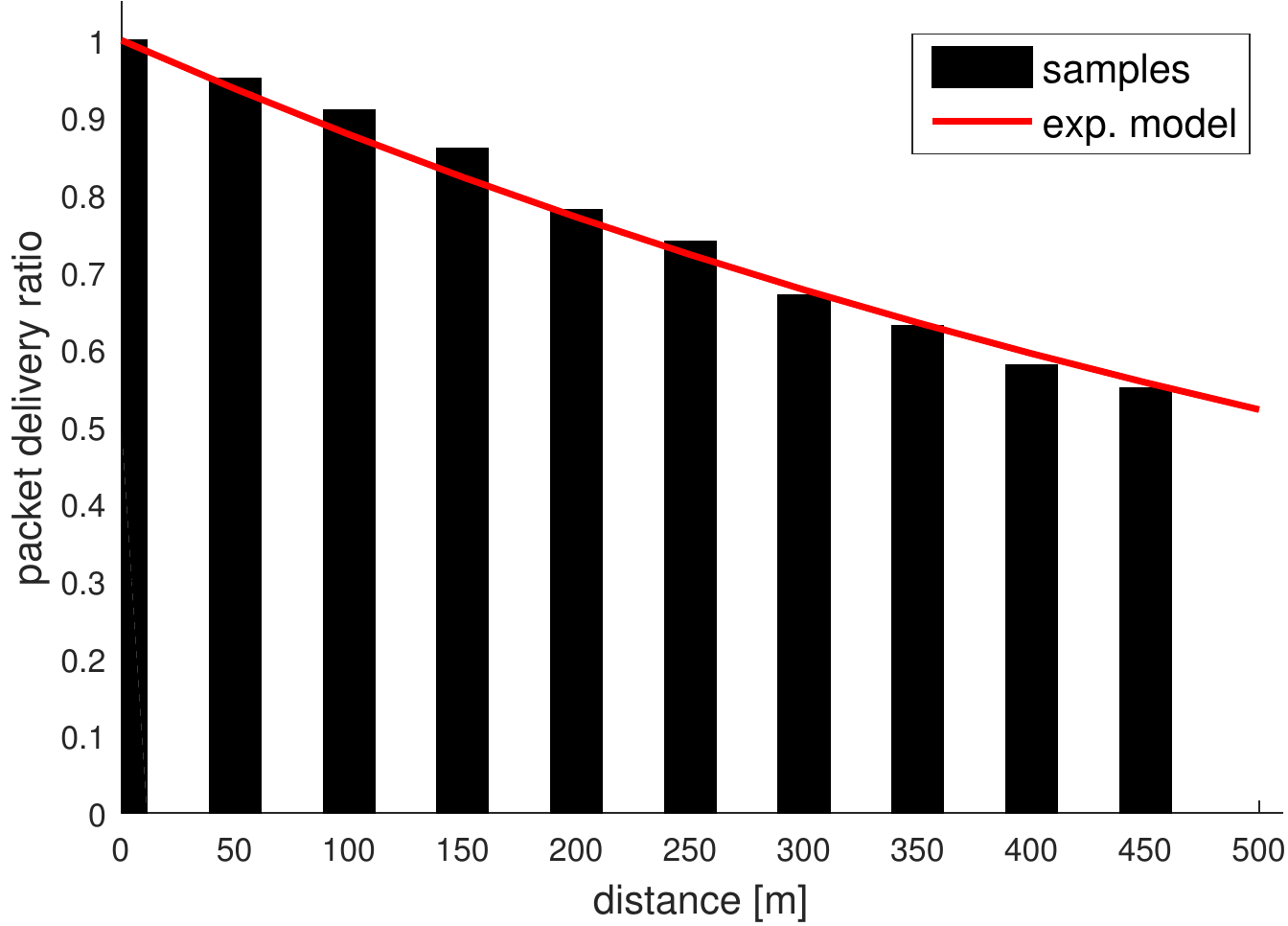}\label{fig:failDistHarsh}}
}
\caption{(a) Packet delivery ratio as function of V2V distance for: (a) open-field, and (b) harsh environment.}
\label{fig:failDist}
\end{figure}

\begin{figure}[!thb]
\centerline{
\includegraphics[width=0.5\textwidth]{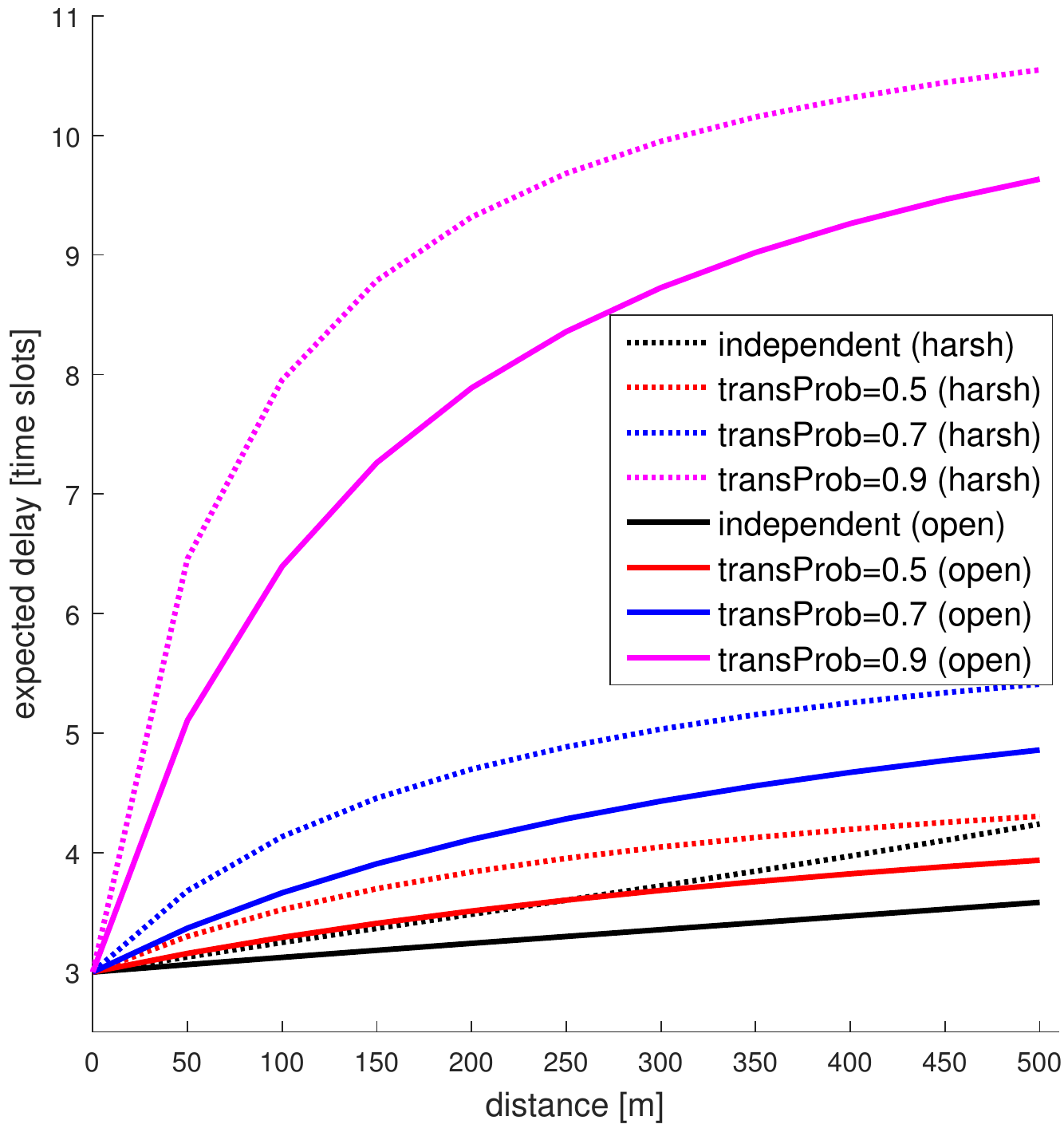}}
\caption{Expected delay as function of V2V distance for open-field and harsh environment, and different transition probabilities.}
\label{fig:delayDistAll}
\end{figure}

\section{Numerical Results}\label{sec:numresults}

Our goal is to analyze the ENTER algorithm caused by V2V failures. For that purpose, we use the real measurements of the packet delivery ratio ($P_{PDR}$), available in \cite{Bai2006}. In this work, authors analyzed 802.11p based DSRC communication for V2V communication, and characterized communication and application level reliability of the wireless channel. They used General Motors (GM) cars equipped with a DSRC radio, omni-directional antenna and a GPS receiver. The transmission power was 20 dBm, the communication range was about 500 m, and the sampling interval was $T=100$ ms. The experiments were conducted on GM test freeways under open-field environment (without any obstacles), and a more realistic harsh environment (with many obstacles, such as tunnels and bridges). 

We use the measurements of $P_{PDR}$ as a function of distance between the cars, for both open-field and harsh environment. Then, we apply exponential model ($e^{-\lambda d}$, where $\lambda$ is the decay rate [$m^{-1}$], and $d$ is the distance [$m$]) to model $P_{PDR}$ as a function of distance. The results are shown in Fig. \ref{fig:failDistOpen} and Fig. \ref{fig:failDistHarsh}, and the corresponding decay rates are, approximately, 0.00063 and 0.0013. 
We then compute the expected delay of the ENTER the algorithm ($\hat{t}_{EN}$) by averaging over different number of communication failures. We consider the scenario in which one car commits $f$ consecutive failures, while other car commits no failures. Assuming independence between time slots, the likelihood of $m$ consecutive failures is given by geometrical distribution: $p(f=m) = (1-P_{PDR})^m \cdot P_{PDR}$, so we can compute the delay as follows:
\be\label{eq:expectedDelay}
\hat{t}_{EN}=\frac{\sum_{m=0}^{F}{p(f=m) \cdot t_{EN}(f=m)}}{\sum_{m=0}^{F}{p(f=m)}}
\ee
We choose $F=30$, which means that cars are allowed to use V2V for up to $(F+1)T=3.1$ s (if $f=F$, both cars would switch to SD after 6.2 s). However, although the independence assumption is experimentally justified in \cite{Bai2006}, it may not be correct if there is a long obstruction of the channel (e.g. due to the large truck in front of the car). For this case, we define a transitional probability $\xi=p(f=m|f=m-1)$ which gives us information about the likelihood of the failure in the current time slot, given that failures already happened in the previous time slot. The likelihood of $m$ consecutive failures is now given by: $p(f=m) = (1-P_{PDR}) \cdot P_{PDR} \cdot \xi^{m-1}$, and then $\hat{t}_{EN}$ can be again computed using \eqref{eq:expectedDelay}.

The results of $\hat{t}_{EN}$ as a function of distance for different values of $\xi$ ($\xi \in \{0.5,0.7,0.9\}$) are shown in Fig. \ref{fig:delayDistAll} for both open-field and harsh environment. As expected, the delay is increasing with distance, and the open-field environment leads to consistently lower delay (although the difference is not significant). However, high transitional probabilities can cause a significant delay, especially for $\xi=0.9$, but this delay is still much lower comparing with the total crossing time that typically takes few seconds. However, in highly unlikely scenario in which $\xi$ is too close to 1, the delay would be too large, so it is crucial that $F$ is carefully chosen to bound this delay.

\begin{figure*}[!t]
\centerline{
\subfloat[]{\includegraphics[width=0.49\textwidth]{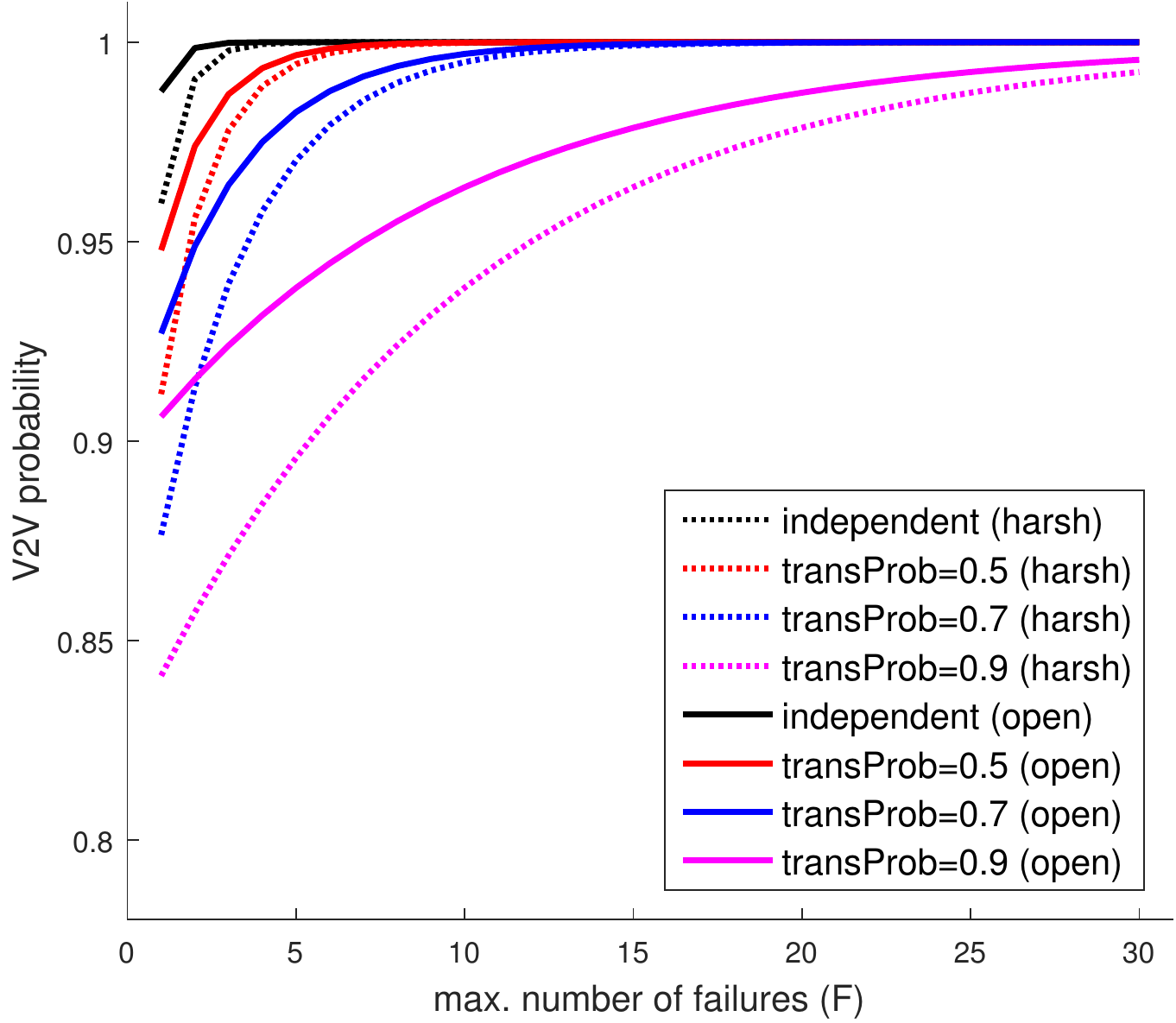}\label{fig:V2Vprob200m}}
\subfloat[]{\includegraphics[width=0.49\textwidth]{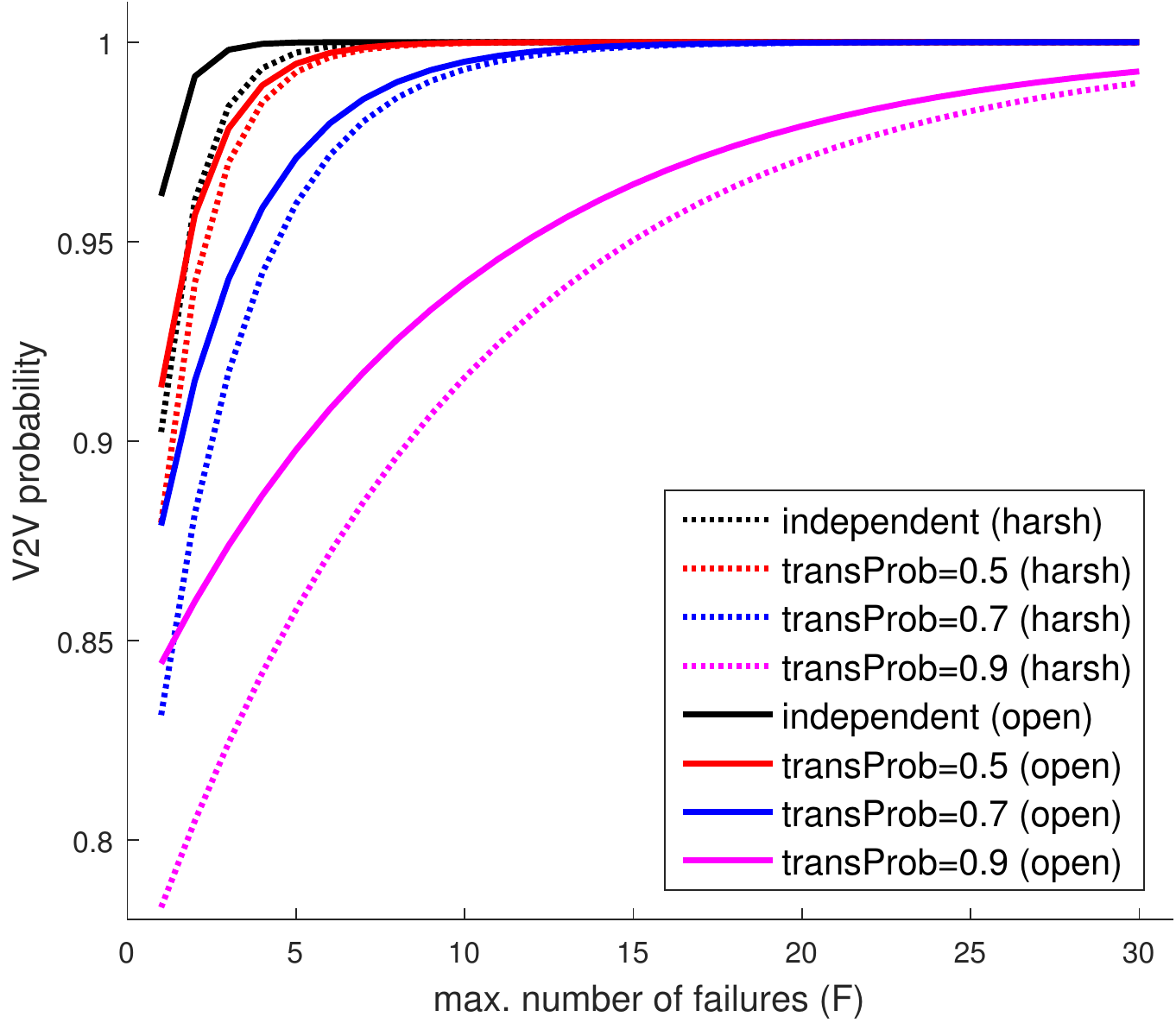}\label{fig:V2Vprob400m}}
}
\caption{V2V probability as function of max. number of failures ($F$) for open-field and harsh environment. V2V distance is (a) 200 m, and (b) 400 m.}
\label{fig:V2Vprob}
\end{figure*}

We then analyse how different parameters influence V2V probability, defined as a percentage of cases in which V2V is used for IC, instead of SD. As we discussed earlier, switch to SD is guaranteed to start after $F+1$ failures, even if these failures happen at just one car. Assuming that there is burst of $F+1$ failures on only one car, V2V probability is given by $P_{V2V} = 1-p(f=F+1)$ and can be computed for both independent and correlated models provided above. The results, for 2 different V2V distances (200 m and 400 m), are shown in Fig. \ref{fig:V2Vprob}. As we can see, the type of environment, max. number of failures ($F$), V2V distance, and transition probabilities influence this probability (the most significant influence is caused by transition probabilities). However, we can make this probability close enough to one by setting $F$ to appropriate value. In this case, $F=15$ (max. delay of 1.5 s) would ensure that V2V is used in at least 95\% of cases. Recall that using V2V, instead of SD, would in most cases lead to much faster IC (especially, when $COL = \emptyset$).

\section{Conclusions and Future Work}\label{sec:conc}

We proposed a novel hybrid IC algorithm, based on both SD and V2V communication. The algorithm is fully distributed, fault-tolerant and uses cars' positions and their dynamics for priority management. Although V2V part of the algorithm can tolerate a limited number of V2V failures, the whole algorithm provides safe and efficient IC even in presence of unlimited number of V2V failures. We provided an analysis of time diagrams that show that both safety and liveness are satisfied in all realistic situations. Moreover, our numerical results showed that the system can be easily configured to use V2V in most scenarios, and that the crossing  delay is just slightly increased even in the presence of correlated failures.

Due to the complexity of the research topic, our results are not complete, so there remain many possibilities for future research. The most interesting question is how much precisely the safety can be improved comparing with SD-based autonomous vehicles from industry (which for now do not use V2V). Another important question is how much the cost of autonomous vehicles would be reduced thanks to V2V, and which sensors can be removed without affecting the safety. There are also other challenging problems, such as development of V2V-based rear-end collision avoidance and platooning algorithms in the presence of V2V failures.

\footnotesize
\bibliographystyle{ieeetr} 
\bibliography{paper-ic-refs}

\end{document}